\documentclass[12pt]{article}
\usepackage{graphicx}
\begin{document}
\setcounter{page}{0}
\thispagestyle{empty}
\begin{flushright}
hep-ph/0602022 \\
\today
\end{flushright}
\begin{center}
{\Large\sc 
Using Invisible Graviton Radiation to Detect Heavy Electroweak Resonances 
at a 500 GeV {\rm e}$^+${\rm e}$^-$ Collider} \\
\vspace*{0.2in}
{\large\sl Sreerup Raychaudhuri {\rm and} Saswati Sarkar} \\
\vspace*{0.3in}
Department of Physics, Indian Institute of Technology, Kanpur 208 016, India. \\
Electronic address: {\sf sreerup@iitk.ac.in, saswati\_05@gmail.com} \\
\vspace*{1.3in}
{\Large\bf ABSTRACT}
\end{center}
\begin{quotation} \noindent \sl
The process $e^+e^- \rightarrow \mu^+ \mu^- \not{\!\!\!E}_T$, where the
missing energy is due to the production of a tower of invisible graviton
states in a model with large extra dimensions, is considered. We focus on
the scenario when this process is used to detect a heavy dileptonic
resonance in the electroweak sector of the model, taking as example some
models with an extra $Z'$ boson. It turns out that at a 500~GeV machine
with 1000~fb$^{-1}$ of luminosity, it may be possible to use this process
to detect such resonances if there are two large extra dimensions and the
string scale is not too far above a TeV.
\end{quotation} \rm\normalsize

\vfill
\newpage

\noindent A concerted international effort is currently under way to build
a high energy linear $e^+e^-$ collider, which, in the wake of expected
discoveries at the LHC, would conduct {\it precision} experiments to probe
the TeV scale.  While discussions on the technical design and feasibility
of such a machine are already under way\cite{ILC}, it is essential, at the
moment, to study the physics possibilities of such a facility and to
determine what would be the best options to have when the experimental
design is finalized. Quite naturally, the study of such physics
possibilities\cite{LCphysics} is heavily dependent on the experience of
the past, where the LEP collider was specifically designed to refine and
complement the discoveries made with the proton synchrotron at CERN. The
difference, in this case, is that we have no unique prediction of the
discoveries expected at the LHC, and hence, the design for a linear
collider must be broad-based and robust enough to adjust to all
contingencies.

\bigskip\noindent Ever since the first serious accelerator experiments in
the 1960s, resonance hunting has always been a major concern of any high
energy experiment. A hadron collider is an ideal machine for seeking
resonances, as the natural variations in parton momenta provide a wide
spread of effective centre-of-mass energies, which then allow the machine
to pick up any unknown resonances within its kinematic reach. This is not
the case at an $e^+e^-$ collider, where the initial state energies are
fixed and have to be decided as part of the machine design. This feature
has obvious advantages for precision measurements, as, for example, at
LEP-1, where the energy was fine-tuned to the $Z^0$ resonance, but there
is always the possibility that there may exist an unknown resonant state
which lies several decay widths away from the actual machine energy, and
consequently, produces a very small effect. A major concern at a linear
$e^+e^-$ collider, would, therefore, be the identification of ways and
means of `seeing' such resonances. This can be achieved if there is a
mechanism to {\it spread out} the initial state energies. Fortunately, a
mechanism is readily at hand, namely initial-state radiation and
beamstrahlung, producing either hard, transverse photons (which can be
tagged) or soft, collinear photons which will go unobserved down the beam
pipe creating a mismatch in the observed momentum balance in the
longitudinal direction\cite{beam}.

\bigskip\noindent In this work, we consider a different -- and somewhat
more exotic -- possibility, namely, that the electroweak physics at a TeV
is embedded in a large-extra-dimensions model of the kind introduced by
Arkani-Hamed, Dimopoulos and Dvali (ADD) in 1998\cite{ADD}.  This model
would then consist of some simple extension (with dileptonic resonances)  
of the Standard Model confined on the `brane' -- a four-dimensional
hypersurface embedded in a 4+$d$ dimensional `bulk' -- with massless
gravitons free to propagate in the bulk. On the brane, where laboratory
experiments must be conducted, the gravitons will appear as multitudes of
massive spin-2 Kaluza-Klein states, each coupling very weakly to matter
and gauge fields, but collectively building up to near-electroweak
strengths in $e^+e^-$ collisions. We focus on the possibility of radiative
processes, where each emitted graviton escapes detection (because of its
extremely weak coupling to matter) and leads to missing energy and
momentum signals. In this paper, to be specific, we have considered the
possibility that an electroweak resonance, like an extra $Z'$ boson, for
example, could be excited in a simple process like $$ e^+e^- \rightarrow
\mu^+ \mu^- \not{\!\!\!E}_T $$ where the missing energy
($\not{\!\!\!E}_T$) due to escaping {\it gravitons} provides the spread in
the effective centre-of-mass energy necessary to home in on the $Z'$
resonance. However, it may be pointed out at this very stage that a
dileptonic resonance need not necessarily be an extra $Z'$ boson -- it
could very well be a scalar, a tensor or an exotic spin state which are
readily obtained in Tev-scale string theories\cite{peskin}.

\bigskip\noindent Irrespective of the specific electroweak model with a
$Z'$ boson (provided the $Z'$ couples to $e^+e^-$ and $\mu^+\mu^-$), there
are seven diagram topologies contributing, at the lowest order, to the
process $e^+e^- \to \mu^+ \mu^- G_n$, with exchange of, respectively, a
photon, a $Z$ boson or a $Z'$ boson, i.e. 21 Feynman diagrams in all.  
These follow from the basic $s$-channel boson-exchange diagram shown in
Fig.~1, with the possibility of a graviton state being radiated, not just
from an initial state (as shown), but from any one of the five internal or
external legs or from any one of the two vertices. One can safely neglect
the corresponding diagrams where a graviton (or a tower of gravitons) is
exchanged in place of the neutral gauge bosons, since the corresponding
amplitudes will be very small\footnote{Suppressed by four extra powers of
$M_S$, the string scale, which will be a TeV or more.}.

\begin{center}
\hskip -10pt
\includegraphics[height=2in]{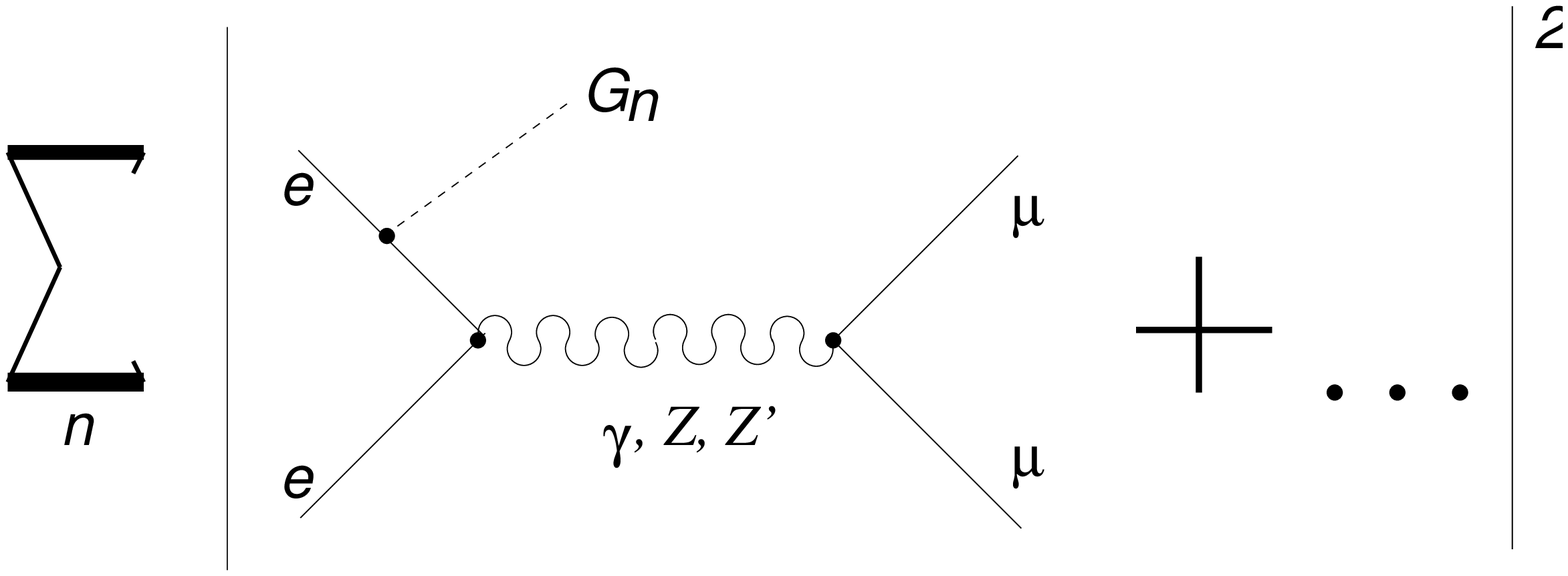}

{\bf Figure 1}. {\footnotesize\it Feynman diagrams contributing to the
process $e^+e^- \to \mu^+ \mu^- G_n$.} \end{center}
\vskip 5pt

\bigskip\noindent It is necessary to make a coherent sum of all 21
diagrams to obtain the cross-section
$$
\sigma_n (M_n) = \sigma (e^+e^- \to \mu^+ \mu^- G_n)
$$
with radiation of a particular graviton Kaluza-Klein mode $G_n$.  The
Feynman rules are given in Refs. \cite{giu_rat_wel, han_lyk_zha}. The
entire calculation, though straightforward, is long and tedious.  We have
performed these calculations by evaluating all the trace relations using
the software FORM\cite{Form}.  Each graviton contributes separately to the
missing energy signal, which means that we require to make an incoherent
sum over all graviton states which are kinematically accessible, i.e.
$$
\sigma = \sum_n \sigma_n (M_n) \simeq \int_0^{\sqrt{s}} dM ~\rho (M) \sigma (M)
$$
making the quasi-continuum approximation for the closely-spaced graviton
states. The density-of-states function $\rho(M)$ is given in
Ref.~\cite{han_lyk_zha}. The actual cross-sections are calculated by
inserting the squared, spin-summed matrix elements into a Monte Carlo
event generator, which provides numerical results of the required degree
of accuracy. In this article we present only the physics results.

\bigskip\noindent In our numerical analysis, we have implemented the
following kinematic cuts:
\begin{enumerate}
\item The final state muons should not be too close to the beam pipe,
which is ensured by demanding that the muon pseudorapidity should satisfy
$ \eta_\mu < 2.5 $ for both the muon tracks. \item The muons should also
have significant transverse momentum, which is ensured by demanding that $
p_T^\mu > 10 \ {\rm GeV} $ for both muon tracks.
\item There should be substantial missing energy, which is ensured by
demanding that $ \not{\!p}_T > 20 \ {\rm GeV} $. It is also useful to
impose an upper cut $ \not{\!p}_T < 200 \ {\rm GeV}$ to get rid of
photon-exchange processes where the muons are very soft.
\item The dimuon invariant mass $M_{\mu\mu} = (p_{\mu^+} + p_{\mu^-})^2$,
which should show a peak at the $Z'$ resonance, should satisfy $
M_{\mu\mu} > 200 \ {\rm GeV} $ which not only removes the peaks due to the
photon and $Z$-exchange effects, but also reduces the SM
background\footnote{Here 'SM background' refers to the SM embedded in an
ADD-type scenario.} substantially. It may be noted that we would not miss
out on resonances in the 100--200~GeV range because these are already
ruled out by LEP-2 data.
\end{enumerate}

\bigskip\noindent For the graviton masses and couplings, we have adopted
the minimal ADD model, which is characterised by two parameters, viz. the
number of compact dimensions $d$ and the (so-called) string scale $M_S$.
The constraints on $M_S$ from purely terrestrial
experiments\cite{ADDbounds} vary between about $0.7$---$1.2$~TeV,
depending on $d$.  We have, in general, chosen $M_S \geq 1$~TeV. In this
paper, in keeping with the general philosophy of laboratory-based
analyses, we choose to ignore the strong astrophysical
bounds\cite{astrobounds} on the $d = 2$ scenario.

\bigskip\noindent There is a much wider choice when it comes to deciding
the parameters of the $Z'$ sector. In general, we can write the couplings
of the $Z'$ boson to fermions ($f$) as
\begin{equation}
{\cal L}^{Z'f\bar f}_{int} = \bar\Psi_f \gamma^\mu \left[ g'^{(n)}_{Lf}P_L
+ g'^{(n)}_{Rf}P_R \right]\Psi_f Z'_\mu
\end{equation}
where $P_{L/R} = \frac{1}{2}(1 \mp \gamma_5)$ and $g'^{(n)}_{Lf}$ and
$g'^{(n)}_{Rf}$. The couplings of the $Z'$ boson to chiral fermions are
different for each scenario ($n$) considered. In order to have a focussed
discussion, we have chosen to follow the pragmatic approach of Dittmar
{\it et al} \cite{dit_nic_djo} which is limited to just five scenarios, of
which the first three are unrealistic and the last two are well motivated.

\begin{enumerate}
\item The couplings of the $Z'$ boson are identical with those of the $Z$
boson; the only difference lies in the mass $M'_Z$. The $\ell^+\ell^-Z'$
vertex (where $\ell = e, \mu$) is given by $g'^{(1)}_{L\ell} \simeq -0.20$
and $g'^{(1)}_{R\ell} \simeq 0.17$.
\item The coupling of the $Z'$ boson is purely vectorlike, with the vector
coupling equal to that of the $Z$ boson; the mass $M'_Z$ is again a free
parameter. The $\ell^+\ell^-Z'$ vertex has $g'^{(2)}_{L\ell} \simeq -0.20$
and $g'^{(2)}_{R\ell} = 0$.
\item The coupling of the $Z'$ boson is purely axial, with the axial
coupling equal to that of the $Z$ boson; as before, the mass $M'_Z$ is a
free parameter. The $\ell^+\ell^-Z'$ vertex has $g'^{(3)}_{L\ell} = 0$ and
$g'^{(3)}_{R\ell} \simeq 0.17$.
\item The $Z'$ boson arises in a $SU(2)_L \times U(1)_Y \times U(1)_{Y'}$
model which originates in the spontaneous breakdown of an $E_6$ group:
\small
\vspace*{-0.7in}
\begin{center} $$
\begin{array}{c}
E_6 \\
\downarrow \\
SO(10) \times U(1)_\psi \nonumber \\
\downarrow \\
SU(5) \times U(1)_\psi \nonumber \\
\downarrow  \\
SU(3)_c \times SU(2)_L \times U(1)_Y \times U(1)_{Y'} \\
\downarrow  \\
SU(3)_c \times SU(2)_L \times U(1)_Y \\
\downarrow  \\
SU(3)_c \times U(1)_{em}
\end{array} $$
\end{center}
\normalsize
In this model, there are two $Z'$-bosons. We focus on the lighter $Z'$,
which is a mixture $ Z' = Z'_\chi \cos \beta + Z'_\psi \sin \beta $, where
$\beta$ varies between $-\pi/2$ to $+\pi/2$, with $\beta = -
\tan^{-1}\sqrt{5/3} \simeq -0.91$ corresponding to direct breaking of
$E_6$ in superstring-inspired models, with no intermediate $SO(10)$ group.  
The $\ell^+\ell^-Z'$ vertex now has $g'^{(4)}_{L\ell} \simeq 0.238 ~\cos
(\beta - 0.406)$ and $g'^{(4)}_{R\ell} \simeq -0.119 ~\sin (\beta
-0.659)$. \item The $Z'$ boson arises in a left-right symmetric model
based on a gauged $SU(2)_L \times SU(2)_R \times U(1)_{B-L}$ symmetry,
with a much simpler breakdown chain
\small
\vspace*{-0.3in}
\begin{center} $$
\begin{array}{c}
SU(3)_c \times SU(2)_L \times SU(2)_R \times U(1)_{B-L} \\
\downarrow  \\
SU(3)_c \times SU(2)_L \times U(1)_Y \\
\downarrow  \\
SU(3)_c \times U(1)_{em}
\end{array} $$
\end{center}
\normalsize
which is presumably the end sector of the breakdown of a bigger symmetry
which unifies colour interactions with the electroweak sector. In this
model, the $\ell^+\ell^-Z'$ vertex has $g'^{(5)}_{L\ell} \simeq
\frac{0.179}{\alpha_{LR}}$ and $g'^{(5)}_{R\ell} \simeq
\frac{0.179}{\alpha_{LR}} ~\left( 1 - \alpha^2_{LR} \right)$. where the
parameter $\alpha_{LR}$ lies between $\sqrt{2/3} \simeq 0.8165$ and
$\sqrt{2} \simeq 1.4142$.
\end{enumerate}

In terms of these parameters, we can write the partial decay width of the
$Z'$ boson to a pair of fermions as
\begin{equation}
\Gamma^{(n)}_{f\bar f} = N_c \frac{M'_{Z}}{24\pi} \sqrt{1 - 4x}
\left[  \left(g'^{(n)}_{Lf}\right)^2 + \left(g'^{(n)}_{Rf}\right)^2
-x \left\{ \left(g'^{(n)}_{Lf}\right)^2 + \left(g'^{(n)}_{Rf}\right)^2
- 6g'^{(n)}_{Lf}g'^{(n)}_{Rf} \right\} \right] \theta(1 - 4x)
\end{equation}
where $x = (m_f/M'_Z)^2$ and $N_c$ is the number of colours of the fermion
$f$. Since, in general, $M'_Z > 100$~GeV from the LEP-2 constraints, we
can neglect $x$ for all but the top quark and get a simplified form
\begin{equation}
\Gamma^{(n)}_{f\bar f} \simeq N_c \frac{M'_{Z}}{24\pi}
\left[  \left(g'^{(n)}_{Lf}\right)^2 + \left(g'^{(n)}_{Rf}\right)^2 \right]
\end{equation}

\bigskip\noindent
The detailed couplings of the $Z'$ to fermions in these five scenarios are
given in Table~1.
\scriptsize
\vspace*{-0.2in}
\begin{center}
$$
\begin{array}{|cc|cc|cc|}
\hline
(n)& (f)& g'^{(n)}_{Lf}
& g'^{(n)}_{Lf}{\rm (num)}& g'^{(n)}_{Rf}& g'^{(n)}_{Rf}{\rm (num)}\\
\hline\hline
(1)            & \nu          & \frac{g}{2c_W} &  ~~0.372 & 0 & 0 \\
               & \ell         &
\frac{g}{c_W}\left( s^2_W -\frac{1}{2} \right) &  -0.201 &
\frac{g}{c_W} s^2_W &  ~~0.171                \\
               & u            &
\frac{g}{2c_W}\left( 1 - \frac{4}{3}s^2_W  \right) &  ~~0.258 &
-\frac{2g}{3c_W}s^2_W &  -0.114 \\
               & d            &              
\frac{g}{3c_W}\left( s^2_W - \frac{3}{2} \right) &  -0.315 &
\frac{g}{3c_W}s^2_W &  ~0.057 \\
\hline
(2) & \nu          & \frac{g}{4c_W} &  ~~0.186 & \frac{g}{4c_W} & ~~0.186 \\
    & \ell         &
\frac{g}{c_W}\left( s^2_W -\frac{1}{4} \right) &  -0.015 &
\frac{g}{c_W}\left( s^2_W -\frac{1}{4} \right) &  -0.015  \\
	                      & u            &
\frac{g}{4c_W}\left( 1 - \frac{8}{3}s^2_W  \right) &  ~~0.072 &
\frac{g}{4c_W}\left( 1 - \frac{8}{3}s^2_W  \right) &  ~~0.072  \\
               & d            &
\frac{g}{3c_W}\left( s^2_W - \frac{3}{2} \right) &  -0.129 &
\frac{g}{3c_W}\left( s^2_W - \frac{3}{2} \right) &  -0.129 \\
\hline
(3) & \nu    & \frac{g}{4c_W} & ~~0.186 & -\frac{g}{4c_W} &  -0.186 \\
    & \ell   & -\frac{g}{4c_W} &  -0.186 & \frac{g}{4c_W} & ~~0.186 \\
    & u      & \frac{g}{4c_W} & ~~0.186 & -\frac{g}{4c_W} &  -0.186 \\
    & d      & -\frac{g}{4c_W} &  -0.186 & \frac{g}{4c_W} & ~~0.186 \\
\hline
(4)            & \nu          & 
\frac{2e}{3c_W} ~\cos \left(\beta - \tan^{-1}\sqrt{\frac{5}{27}}\right) & 
 0.238 ~\cos (\beta - 0.406) &  0 & 0 \\
               & \ell         & 
\frac{2e}{3c_W} ~\cos \left(\beta - \tan^{-1}\sqrt{\frac{5}{27}}\right) &
 0.238 ~\cos (\beta - 0.406) & 
-\frac{e}{3c_W} \sin \left(\beta - \tan^{-1}\sqrt{\frac{3}{5}}\right) &
-0.119 ~\sin (\beta -0.659) \\
               & u            & 
\frac{e}{3c_W} ~\sin \left(\beta - \tan^{-1}\sqrt{\frac{3}{5}}\right) &
 0.119~\sin (\beta - 0.659) &
-\frac{e}{3c_W} ~\sin \left(\beta - \tan^{-1}\sqrt{\frac{3}{5}}\right) &  
-0.119 ~\sin (\beta - 0.659) \\
               & d            & 
\frac{e}{3c_W} ~\sin \left(\beta - \tan^{-1}\sqrt{\frac{3}{5}}\right) &
0.119~\sin (\beta - 0.659) &
-\frac{2e}{3c_W} ~\cos \left(\beta - \tan^{-1}\sqrt{\frac{5}{27}}\right) &
-0.238 ~\cos (\beta - 0.406) \\
\hline
(5)            & \nu          & 
\frac{e}{2c_W}\frac{1}{\alpha_{LR}} &  \frac{0.1785}{\alpha_{LR}} &  0 & 0 \\
               & \ell         &
\frac{e}{2c_W}\frac{1}{\alpha_{LR}} &  \frac{0.1785}{\alpha_{LR}} &
\frac{e}{2c_W}\frac{1}{\alpha_{LR}} \left( 1 - \alpha^2_{LR} \right)
&  \frac{0.1785}{\alpha_{LR}} \left( 1 - \alpha^2_{LR} \right)  \\
               & u            &
-\frac{e}{6c_W}\frac{1}{\alpha_{LR}} &  -\frac{0.0595}{\alpha_{LR}} &
-\frac{e}{6c_W}\frac{1}{\alpha_{LR}} \left( 1 - 3\alpha^2_{LR} \right) 
&  -\frac{0.0595}{\alpha_{LR}} \left( 1 - 3\alpha^2_{LR} \right) \\
               & d            &
-\frac{e}{6c_W}\frac{1}{\alpha_{LR}} &  -\frac{0.0595}{\alpha_{LR}} &
-\frac{e}{6c_W}\frac{1}{\alpha_{LR}} \left( 1 + 3\alpha^2_{LR} \right) 
&  -\frac{0.0595}{\alpha_{LR}} \left( 1 + 3\alpha^2_{LR} \right) \\
\hline\hline
\end{array}
$$
\end{center}
\normalsize
{\bf Table 1.} {\it Couplings of the $Z'$ boson to different fermions in
the five models under consideration. We write $c_W = \cos \theta_W$ and
$s_W = \sin \theta_W$.}
\bigskip

\bigskip\noindent It is now vital to numerically evaluate the total decay
width of the $Z'$ boson in these models and determine if, indeed, it will
appear as a narrow resonance. This is illustrated in Figure~2, where we
have plotted the total decay width $\Gamma_{Z'}$ as a function of $M'_Z$
for different values of the couplings in the two realistic models in
question, i.e. (4) and (5). The hatched area corresponds to the $E_6$
model, i.e. (4), while the dotted area corresponds to the left-right
symmetric model, i.e. (5). These areas have been generated by varying the
couplings $\beta$ and $\alpha_{LR}$ over the full allowed range (see
above).  The three solid lines show the width in the (unrealistic)
comparison models (1), (2) and (3).  Small deviations from linearity in
all the curves correspond to opening-up of the $t\bar t$ decay channel.
The dark-shaded area is ruled out by the direct LEP-2 constraints.

\begin{center}
\hskip -10pt
\includegraphics[height=3.5in]{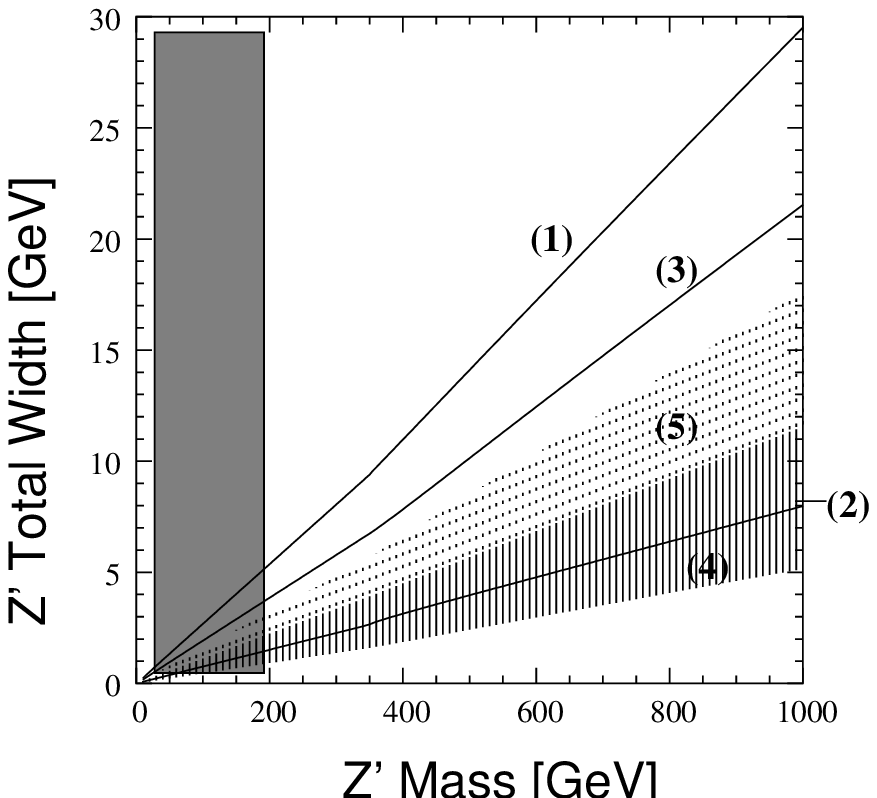}

\end{center}
{\bf Figure 2}.
{\footnotesize\it
Decay width of the $Z'$ boson in five different models as a function of
its mass. Numbers in parentheses correspond to the number of the model as
described in the text. The dark-shaded region is ruled out by LEP-2.} 
\vskip 5pt

It may be seen that in the two realistic models (4) and (5), the decay
width of the $Z'$ boson does not rise above 17~GeV, even when the mass
reaches 1~TeV. Since this indicates $\Gamma_{Z'}/M_{Z'} < 2$\%, we may
safely conclude that the $Z'$-boson will appear as a narrow resonance in
the discussion that follows.

\bigskip\noindent In any of the above models, the probability of exciting
a $Z'$ resonance in the process $e^+e^- \to \mu^+ \mu^- G_n$ will depend
on the probability of emitting initial-state gravitons carrying away the
requisite energy from the electron-positron pair. This means that the size
of the resonant cross-section will depend rather sensitively on the ADD
parameters, viz., $d$, the number of compact dimensions, and $M_S$, the
string scale. This is illustrated in Figure 3, where we plot the total
cross-section as a function of $M_S$, for a fixed set of all the other
parameters, mostly chosen to get a high cross-section. Models~2 and 3
yield cross-sections in the same ballpark as Model~1, and have been
omitted to avoid making the figure clumsy.

\begin{center}
\hskip -10pt
\includegraphics[height=3.5in]{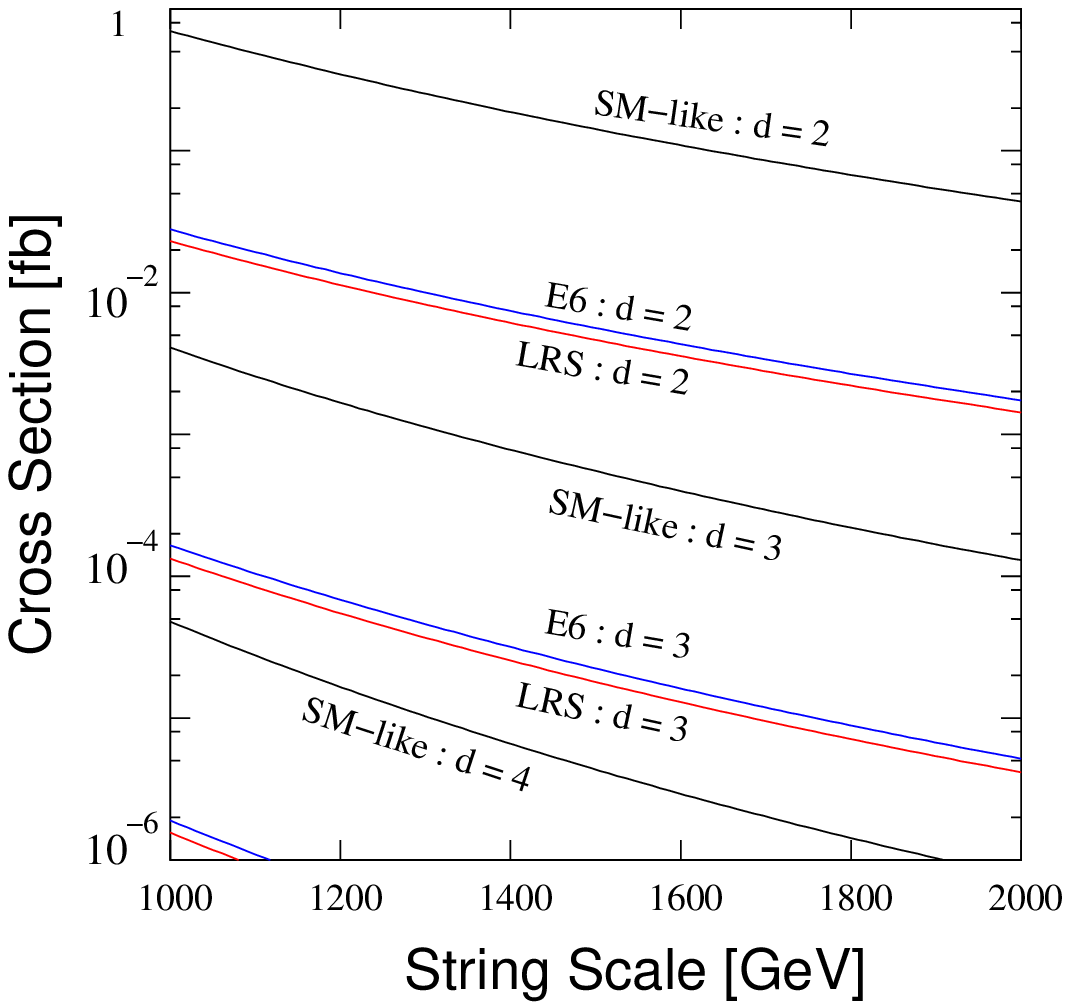}
                                                                                
\end{center}
{\bf Figure 3}.
{\footnotesize\it Cross-section for $e^+e^- \rightarrow \mu^+ \mu^-
\not{\!\!\!E}_T$ as a function of $M_S$, keeping $d = 2$ and $M_{Z'} =
250$~GeV. The black curves correspond to Model 1, i.e. the couplings of
the $Z'$ boson to fermions are identical to the couplings of the $Z$
boson. Blue curves correspond to the $E_6$-based Model~4 with $\beta = 0$,
the last being chosen to correspond to large cross-section. Red curves
correspond to a LR-symmetric Model~5 with $\alpha_{LR} \simeq \sqrt{2/3}$,
which is also chosen to get the largest cross-section. }
\vskip 5pt

\bigskip\noindent It immediately becomes clear from Figure~3 that the
present process is viable at a 500~GeV collider, with an integrated
luminosity of around 1000~fb$^{-1}$, only if we assume that $d = 2$ and
$M_S$ is not much greater than 1.8~GeV. For $d = 3$, one would require a
much higher luminosity of at least $10^6$~fb$^{-1}$ in order to have
observable effects. Higher values of $d$ are simply not accessible at the
planned energies and luminosities.
  
\bigskip\noindent The above curves correspond to a $Z'$-boson mass of
250~GeV. For the same choice of mass, and with $d = 2$ and $M_S = 1$~TeV,
we have plotted, in Figure~4, the distribution in dimuon invariant mass $
M_{\mu\mu} = \left( p_{\mu^+} - p_{\mu^-} \right)^2 $ for all the five
models in question.

\begin{center}
\hskip -10pt
\includegraphics[height=6.5in]{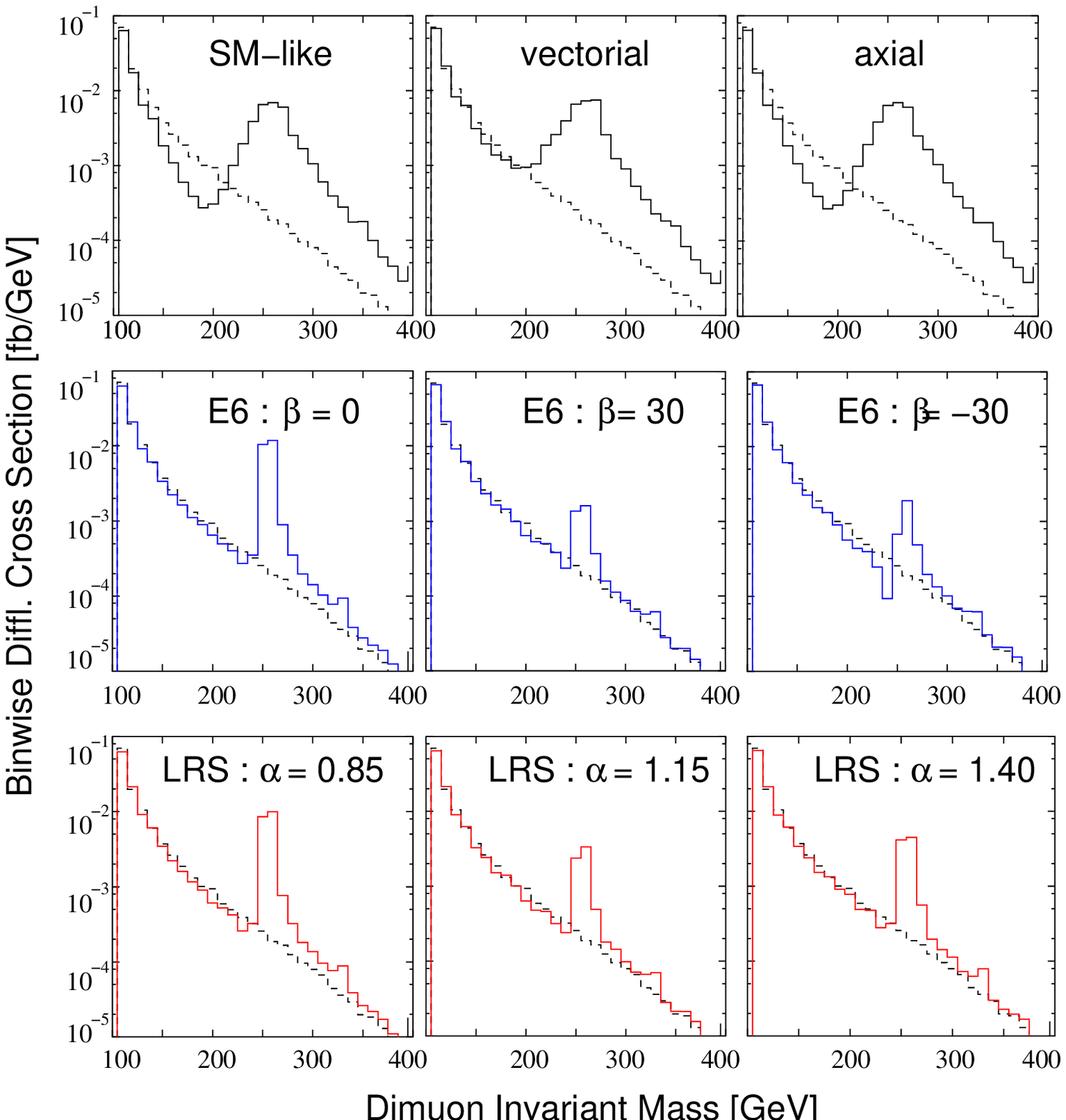}
                                                                                
\end{center}
{\bf Figure 4}.
{\footnotesize\it Bin-wise distribution of the differential cross-section
in dimuon invariant mass $M_{\mu\mu}$ in different models. The dashed
histogram corresponds, in every case, to absence of any $Z'$ resonance,
i.e. to a SM on the brane.  The top row illustrates the cases for Models
1, 2 and 3. The middle row illustrates Model~4 for different choices of
mixing angle $\beta$, while the bottom row similarly illustrates the case
of different choices of $\alpha_{LR}$. }
\vskip 5pt

\bigskip\noindent In Figure~4, the dashed histogram represents the case
when there is no $Z'$ resonance in the given range, i.e. one observes only
the SM on the brane. This falls off sharply as $M_{\mu\mu}$ increases
because of the usual $s$-channel suppression far away from the photon and
$Z$ resonances. The presence of a $Z'$ boson changes this behaviour
dramatically, as shown by the solid histograms. The first three,
corresponding to Models 1, 2 and 3 respectively, have large and wide
resonances, consistent with the larger width and couplings illustrated in
Table~1 and Figure~2. The behaviour of `SM-like' and `axial' couplings for
the $Z'$ boson is very similar simply because the vector coupling of
charged leptons to the $Z$ boson is very small in the SM. There is a
qualitative difference between these and a purely vectorial coupling, in
that the latter does not show any destructive interference effects. More
realistic results are shown in the blue and red histograms. The blue
histograms correspond to the $E_6$ model with choices of $\beta = 0,
\pi/6$ and $-\pi/6$ respectively. Though the model allows $\beta$ to vary
from $-\pi/2$ to $+pi/2$, we found that the qualitative behaviour of the
invariant mass distribution shows very little variation beyond $\beta =
\pm \pi/6$. It is quite clear that $\beta = 0$ provides the largest
resonances. For finite values of $\beta$ the resonances look small in the
figure, but are not actually so, since the ordinate is plotted on a
logarithmic scale. Similar arguments hold for the red histograms in the
bottom row, which correspond to Model 5, with different values of the
coupling constant $\alpha_{LR} = 0.85, 1.15$ and 1.40 respectively,
spanning the allowed range. As expected from the $1/\alpha_{LR}$
dependence shown in Table~1, the largest resonances correspond to the
smallest value of $\alpha_{LR}$.

\bigskip\noindent We reiterate at this point that the actual point in the
parameter space of the model incorporating $Z'$ bosons is a matter of
detail and our work does not focus on this. Figure~4 simply illustrates
the fact that with quite realistic choices of parameters in this sector,
we can expect large resonances in the dimuon mass distribution in the
process $e^+e^- \rightarrow \mu^+ \mu^- \not{\!\!\!E}_T$, provided the
model is embedded in an ADD-like brane world scenario.

\bigskip\noindent In all of the above analyses, the mass of the $Z'$ boson
was set to 250~GeV. It is natural to ask whether it would be possible to
distinguish resonances when $M_{Z'}$ is larger and approaches the machine
energy of 500~GeV. Obviously, in this case one would get smaller and
broader resonances, which may not show such spectacular deviations from
the SM expectations as shown in Figure~4. In this case, it is better to
have a quantitative index, for which we have adopted a $\chi^2$ fit to the
line-shape. This has been done as follows: the differential cross-section
has been calculated in 14 bins of 20~GeV each, starting from $M_{\mu\mu} =
200$~GeV to $M_{\mu\mu}$ = 480~GeV. Denoting the cross-section in the
$i$th bin by $\sigma_i$ and choosing a luminosity $L = 10^3$~fb$^{-1}$, we
then calculate the number of events expected in each bin as $N_i =
L\sigma_i$. We separately calculate the expected number for $N_i^{(SM)}$
and $N_i^{Z'}$, i.e. without and with the $Z'$ resonance. If the expected
number drops below unity, we set it to unity, to take care of random
fluctuations and the fact that the law of large numbers is clearly not
valid. The $\chi^2$ is now calculated, assuming Gaussian errors
(statistical only) $\delta N_i^{(SM)} = \sqrt{N_i^{(SM)}}$, using the
simple formula
\begin{equation}
\chi^2 = \sum_{i=1}^{14} 
\frac{\left( N_i^{Z'} - N_i^{(SM)} \right)^2 } { N_i^{(SM)} } 
\end{equation}
The value of this $\chi^2$ is a measure of the deviation of the line-shape
in $M_{\mu\mu}$ from the SM expectations and consequently falls as the
resonances become smaller. For $\chi^2 < 23.7$, the line-shape is
consistent with Gaussian random fluctuations in the SM at 95\% confidence
level (C.L.), which means that this value of $\chi^2$ corresponds to the
95\% discovery limit of the resonance.

\begin{center}
\hskip -10pt
\includegraphics[height=3.5in]{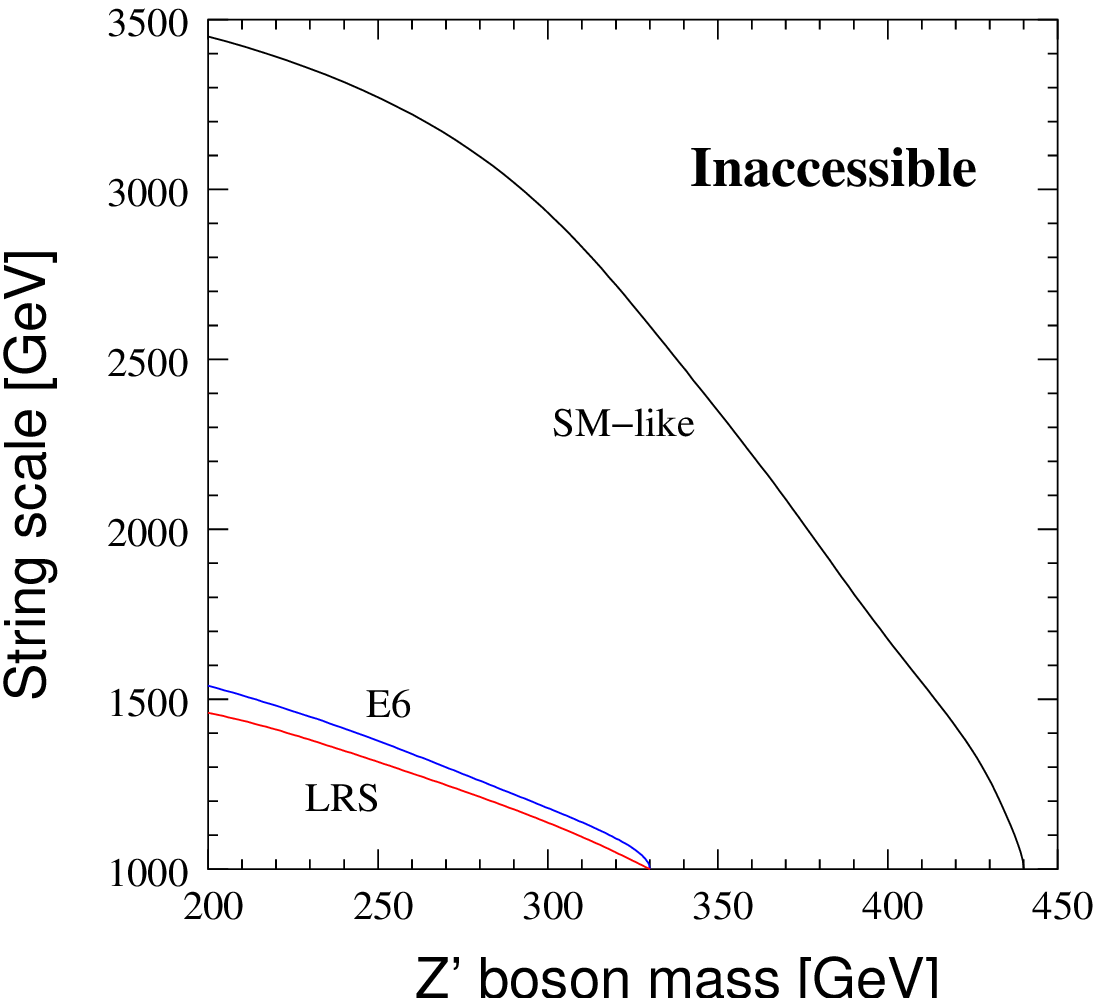}
                                                                                
\end{center}
{\bf Figure 5}.
{\footnotesize\it 95\% C.L. discovery limits in the $M_S$--$M_{Z'}$ plane
corresponding to three different models. The black curve corresponds to a
$Z'$ boson with SM-like couplings to fermions. The blue curve corresponds
to an $E_6$-based model with $\beta = 0$ and the red curve corresponds to
a LR symmetric model with $\alpha_{LR} \simeq \sqrt{2/3}$.  }
\vskip 5pt

In Figure~5, we have plotted these 95\% C.L. discovery limits in the
$M_S$--$M_{Z'}$ plane, the region {\it below} the curve being accessible
to a 500~GeV linear collider with $10^3$~fb$^{-1}$ luminosity. The three
curves all correspond to $d = 2$ and, respectively, Model~1 (black),
Model~4 (blue) with $\beta = 0$ and Model~5 (red) with $\alpha_{LR} =
0.85$, these being (as we have seen) the choices which lead to the largest
cross-sections. As expected from the large cross-sections in Figure~3, the
discovery limits are the most robust in the (unphysical) case of Model~1,
and the ones for the more realistic cases look disappointing by contrast.
Nevertheless, it should be possible, for a string scale not much larger
than a TeV, to detect $Z'$ resonances all the way up to nearly 330~GeV.
Given the kinematic constraints at a machine energy of 500~GeV, this is
quite a reasonable effect. Moreover, as stated above, the purpose of this
work is to highlight the possibility of exciting resonances of perhaps
unknown nature, and not to focus on the details of models with $Z'$
bosons.

\bigskip\noindent In this paper, therefore, we have investigated the
possibility that, at a 500~GeV $e^+e^-$ collider, invisible graviton
radiation in a brane world scenario could provide a useful tool to
discover low-lying resonances coupling to $e^+e^-$ and $\mu^+\mu^-$ pairs.
Not only are such resonances predicted in extensions of the standard
electroweak model, but some might be low-lying stringy excitations arising
in a theory with TeV strings. We have explored the collider phenomenology
of this process by focussing on extra $Z'$ bosons and showed that
reasonably optimistic discovery limits are predicted in two of the more
popular models. In so doing, we have also established the methodology for
investigation of this particular signal. It may be noted that at a 500~GeV
$e^+e^-$ collider, one of the easiest processes to look for will be a hard
muon pair with substantial missing energy, and we shall surely have data
on this when the machine actually runs. The possibility that a new
resonant state will be discovered --- or, at least, confirmed --- in this
process is by no means a far-fetched one, and this is the substance of the
present work. Quite obviously, there are severe limitations due to the
limited energy and luminosity available at the machine in question, and
one could carry out a more extensive study if, for example, the collisions
took place at $\sqrt{s} = 2$~TeV and an integrated luminosity of
$10^6$~fb$^{-1}$ could be achieved\cite{CLIC}.  However, we feel that it
is premature to carry out any numerical analysis based on such machine
parameters. Our point is sufficiently conveyed by the rather spectacular
resonant structures shown in, for example, Figure~4, and a study of the
remaining aspects could await the development of a detailed machine design
for the International Linear Collider.

\bigskip\bigskip\noindent {\small The authors acknowledge useful
discussions with S.K.~Rai. SR acknowledges the hospitality of the
Institute of Physics, Bhubaneswar, India, where a part of this work was
done. The work of SS was funded by the Department of Science and
Technology, Government of India.}

\newpage


\begin{thebibliography}{99}
\bibitem{ILC} For a recent status report, see, for example, K.~Moenig,
{\it Acta Phys.Polon.} {\bf B36}, 3327 (2005).
\bibitem{LCphysics} See, for example, E.~Accomando {\it et al.}, {\it
Phys.Rept.} {\bf 299}, 1 (1998).
\bibitem{beam} T.~Buanes, E.W.~Dvergsnes and P.~Osland, {\it Eur.Phys.J.}
{\bf C35}, 555 (2004); \\ N.K.~Mondal et al., {\it Pramana} {\bf 63}, 1331
(2004).
\bibitem{ADD} N.~Arkani-Hamed, S.~Dimopoulos and G.R.~Dvali, {\it
Phys.Lett.} {\bf B429}, 263 (1998) and {\it Phys.Rev.} {\bf D59}, 086004
(1999);\\ I.~Antoniadis, N.~Arkani-Hamed, S.~Dimopoulos and G.R.~Dvali,
{\it Phys.Lett.} {\bf B436}, 257 (1998).
\bibitem{peskin} S.~Cullen, M.~Perelstein and M.E.~Peskin, {\it Phys.Rev.}
{\bf D62}, 055012 (2000).
\bibitem{giu_rat_wel} G.F.~Giudice, R.~Rattazzi and J.D.~Wells, {\it
Nucl.Phys.} {\bf B544}, 3 (1999).
\bibitem{han_lyk_zha} T.~Han, J.D.~Lykken and R.-J.~Zhang, {\it Phys.Rev.}
{\bf D59}, 105006 (1999).
\bibitem{Form} See, for example, J.A.M.~Vermaseren, math-ph/0010025.
\bibitem{ADDbounds} See, for example, Y.~Uehara, {\it Mod.Phys.Lett.} {\bf
A17}, 1551 (2002).
\bibitem{astrobounds} S.~Cullen and M.~Perelstein, {\it Phys.Rev.Lett.}
{\bf 83}, 268 (1999); \\ V.D.~Barger {\it et al.} {\it Phys.Lett.} {\bf
B461}, 34 (1999).
\bibitem{dit_nic_djo} M.~Dittmar, A.-S.~Nicollerat and A.~Djouadi, {\it
Phys.Lett.} {\bf B583}, 111 (2004).
\bibitem{CLIC} See, for example, the report by the CLIC Physics Working
Group (E.~Accomando {\it et al.}), hep-ph/0412251.
\end{thebibliography}
\end{document}